\documentclass[9pt,twocolumn,twoside]{opticajnl}
\journal{optica} 
\usepackage{stfloats}
\setboolean{shortarticle}{false}

\usepackage{lineno}

\title{Inverse design of coherent supercontinuum generation using free-form nanophotonic waveguides}

\author[1,$\dagger$]{Chia-Yi Lee}
\author[1,$\dagger$]{Yanwu Liu}
\author[1]{Yinke Cheng}
\author[1]{Chenghao Lao}
\author[1,2,3]{Qihuang Gong}
\author[1,2,3*]{Qi-Fan Yang}

\affil[1]{State Key Laboratory for Artificial Microstructure and Mesoscopic Physics and Frontiers Science Center for Nano-optoelectronics, School of Physics, Peking University, Beijing 100871, China}
\affil[2]{Collaborative Innovation Center of Extreme Optics, Shanxi University, Taiyuan, China}
\affil[3]{Peking University Yangtze Delta Institute of Optoelectronics, Nantong, Jiangsu, 226010, China.}
\affil[$\dagger$]{These authors contributed equally to this work.}
\affil[*]{leonardoyoung@pku.edu.cn}

\begin{abstract}
Many key functionalities of optical frequency combs such as self-referencing and broad spectral access rely on coherent supercontinuum generation (SCG). While nanophotonic waveguides have emerged as a compact and power-efficient platform for SCG, their geometric degrees of freedom have not been fully utilized due to the underlying nonlinear and stochastic physics. Here, we introduce inverse design to unlock free-form waveguides for coherent SCG. The efficacy of our design is numerically and experimentally demonstrated on Si$_3$N$_4$ waveguides, producing flat and coherent spectra from visible to mid-infrared wavelengths. Our work has direct applications in developing chip-based broadband light sources for spectroscopy, metrology, and sensing across multiple spectral regimes.
\end{abstract}

\setboolean{displaycopyright}{false} 

\begin{document}

\date{}
\maketitle
\section{Introduction}

Formed by equally-spaced spectral lines with fixed phase relations, optical frequency combs (OFCs) have received recent intense interest for their revolutionary impacts on timing, sensing, and spectroscopy \cite{diddams2020optical}. These diverse applications demand OFCs operated at different wavelengths, from visible bands for astronomical calibration \cite{Murphy2007} to mid-infrared (mid-IR) bands for trace-gas detection \cite{schliesser2012mid}. Such spectral windows can be unlocked through supercontinuum generation (SCG), which spectrally broadens narrow-band comb generators via third-order nonlinear optical processes. Indeed, SCG has become an essential component in many OFCs to realize octave-spanning spectra for $f-2f$ self-referencing \cite{telle1999carrier}. Over the past decade, the platforms for SCG have evolved from microstructured fibers \cite{dudley2006supercontinuum} to nanophotonic waveguides \cite{gaeta2019photonic}, with increased optical confinement and Kerr nonlinearity. A wide selection of waveguide materials such as Si \cite{hsieh2007supercontinuum, kou2018mid}, SiO$_2$ \cite{yoon2017coherent}, Si$_3$N$_4$ \cite{halir2012ultrabroadband, guo2018mid}, LiNbO$_3$ \cite{yu2019coherent, lu2019octave}, AlN \cite{hickstein2017ultrabroadband,lu2020ultraviolet,chen2021supercontinuum}, Ta$_2$O$_5$ \cite{fan2019visible,woods2020supercontinuum,fan2021higher}, and AlGaAs \cite{kuyken2020octave, may2021supercontinuum} have provided a large transparency window for multi-octave SCG from ultra-violet to mid-infrared wavelengths \cite{lu2020ultraviolet}.

In addition to the spectral broadening, the parasitic amplification of noises during SCG could destroy the comb structure. Despite the established theoretical model, designing coherent SCG is usually empirical (e.g., using short pulses and short waveguides) due to the underlying nonlinear and stochastic processes \cite{agrawal2000nonlinear,dudley2006supercontinuum}. Recently, inverse design has shown the potential to expedite SCG design by generating waveguide geometries from target spectra \cite{molesky2018inverse,genty2021machine}. However, incorporating coherence as an objective for optimization remains challenging and time-consuming since most inverse-design algorithms are not compatible with stochastic processes. As a result, geometric optimization of waveguides \cite{Ciret:17,wei2020supercontinuum} and fibers \cite{zhang2009genetic} for coherent SCG have only been tested with few degrees of freedom.

\begin{figure*}
\centering
\includegraphics[width=\linewidth]{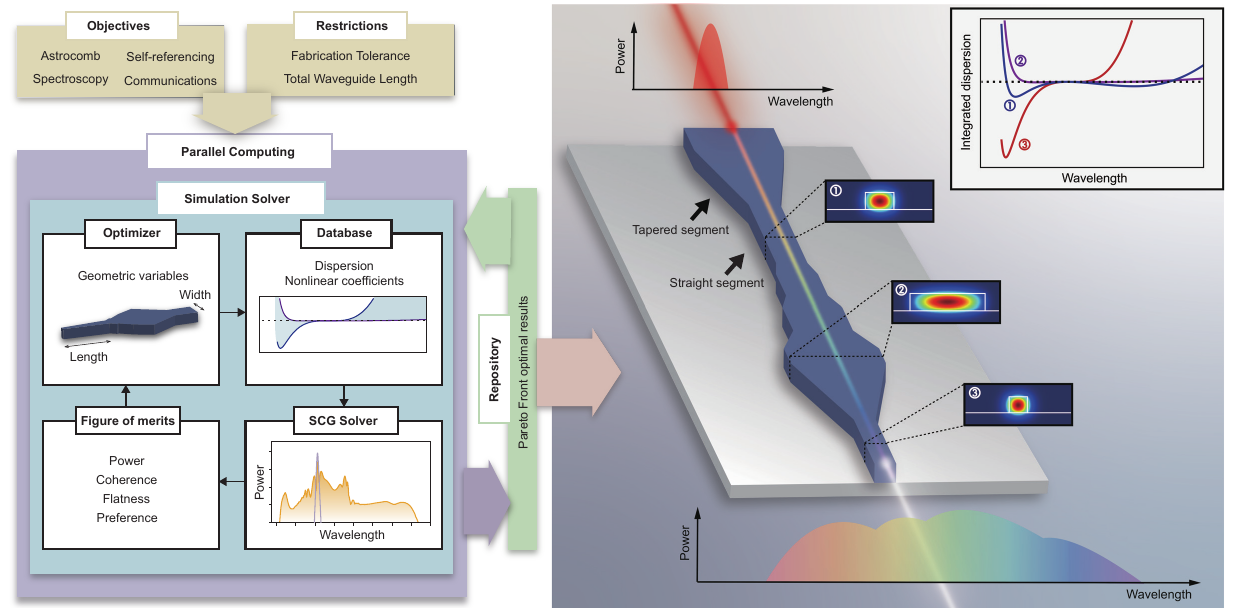}
\caption{{\bf Working flow of the inverse design algorithm.} The algorithm contains four main parts namely objectives, restrictions, simulation solver, and repository. The right panel shows the geometry of the free-form waveguide, with its cross-sectional modal profiles and dispersion profiles varying at different locations.}
\label{Fig1}
\end{figure*}

In this work, we introduce an efficient protocol for the inverse design of coherent SCG. The waveguide widths along the propagation direction are freely varied, which impacts both the local dispersion and nonlinear coefficient. Utilizing the Particle Swarm Optimization (PSO) algorithm, multiple objectives, including power, spectral flatness, and coherence, can be simultaneously optimized according to preset preference levels. The advantages of the inverse-designed free-form waveguides over uniform waveguides are numerically and experimentally verified on $\rm Si_{3}N_{4}$ platforms. Additionally, an automated algorithm is developed for the optimal strategy to create self-referenced OFCs.

\section{Methodology}

The waveguide geometry and computational flow are depicted in Fig. \ref{Fig1}. The free-form waveguide comprises multiple straight segments with distinct lengths and widths, which are connected by tapered segments for adiabatic transitions. To compute the dispersion and effective mode area for a given cross-section, a vector finite difference mode solver is employed \cite{fallahkhair2008vector}. To expedite the computations, a database is established to store waveguide properties for subsequent simulations.

The SCG solver is based on the generalized nonlinear Schr\"{o}dinger equation (GNLSE) \cite{laegsgaard2007mode,dudley2010supercontinuum}:
\begin{equation}\label{Eq:4}
\begin{split}
  &\frac{\partial \left(\tilde{A}(z, \omega) \exp (-\hat{D}(\omega) z)\right)}{\partial z} \\
  &= i \frac{n_{2} n_{0}(\omega) \omega}{c n_{\mathrm{eff}}(\omega) A_{\mathrm{eff}}^{1 / 4}(\omega)} \exp (-\hat{D}(\omega) z)\mathcal{F}\left\{\bar{A}(z, T)|\bar{A}(z, T)|^2\right\},
\end{split}
\end{equation}
The dispersion operator \(\hat{D}(\omega)\) and the function \(\bar{A}(z, T)\) are defined as follows:
\begin{equation}\label{Eq:5}
\bar{A}(z, T)=\mathcal{F}^{-1}\left\{\frac{\tilde{A}(z, \omega)}{A_{\mathrm{eff}}^{1 / 4}(\omega)}\right\}, 
\end{equation}
\begin{equation}\label{Eq:6}
\hat{D}(\omega)=i\left(\beta(\omega)-\beta\left(\omega_{0}\right)-\beta_{1}\left(\omega_{0}\right)\left(\omega-\omega_{0}\right)\right)-\alpha(\omega) / 2.
\end{equation}
where $\tilde{A}(z, \omega)$ represents the complex spectral envelope, $\alpha(\omega)$ denotes the linear loss, $\beta_{1}$ corresponds to the first-order dispersion, $A_{\mathrm{eff}}$ is the effective mode area, \(n_{2}\) refers to the nonlinear index, \(n_{0}\) represents the linear refractive index, \(n_{\mathrm{eff}}\) signifies the effective index, and $c$ is the speed of light in vacuum. Using data retrieved from the database, the SCG solver generates an output spectrum in approximately one minute.

We present the inverse design process as an optimization problem, focusing on three figures of merit ($FOM$s): power, flatness, and coherence within a specific spectral window ($\omega_{1}$ to $\omega_{2}$). The $FOM$s for power and flatness are represented by the following equations:
\begin{equation}\label{eq:1}
FOM_{1}=-\int_{\omega_{1{\ }}}^{\omega_{2{\ }}}{c(\omega)\log\left(|\tilde{A}\left(\omega\right)\right|^2)\mathrm{\ d}\omega},\\
\end{equation}
\begin{equation}\label{eq:2}
FOM_{2}=-\int_{\omega_{1{\ }}}^{\omega_{2{\ }}}\frac{\left[\log \left(|\tilde{A}(\omega)|^2\right)-\overline{\log\left(|\tilde{A}(\omega)|^2\right)}\right]^2}{\overline{\log\left(|\tilde{A}(\omega)|^2\right)}^2(\omega_{2{\ }}-\omega_{1})} \mathrm{~d} \omega,
\end{equation}
Here, $c(\omega)$ represents a weight function. To avoid the generation of large spikes on the spectrum, we use the logarithm of the power spectral density ($\log\left(|\tilde{A}(\omega)|^2\right)$) rather than the absolution power levels. The coherence is evaluated by simply comparing the difference between two realizations ($\tilde{A}_1$ and $\tilde{A}_2$) that are independently simulated with and without noise seedings added to the input pulse:
\begin{equation}\label{eq:3}
FOM_{3}=-\int_{\omega_{1{\ }}}^{\omega_{2{\ }}}\left[\log{\left(|\tilde{A}_1\left(\omega\right)\right|^2)}-\log{\left(|\tilde{A}_2\left(\omega\right)\right|^2)}\right]^2\mathrm{d} \omega.
\end{equation}
We find that $FOM_3$ is sufficient for the optimization process, which is further verified using coherence calculated from more independent realizations (see section 3).

To minimize these $FOM$s simultaneously, we employ the PSO algorithm \cite{kennedy2006swarm} and store the optimal solutions in a repository using Pareto dominance \cite{coello2004handling}. $FOM$ preference guidance is introduced to ensure that the population evolves towards $FOM$s with higher priority in each iteration, thus avoiding unnecessary solutions on the edges of the solution set. Constraints, such as the total length of the waveguide and minimum feature size during fabrication, are also included. Finally, the solutions in the repository are refined to reveal the optimal results. In contrast to genetic algorithms \cite{zhang2009genetic,Ciret:17}, the PSO algorithm is insensitive to initial values. The numerous local minima of the GNLSE can be avoided since the PSO actively exchanges local and global information. Notably, PSO is compatible with parallel computation which can expedite the optimization. For example, on a regular desktop PC, five simulation solvers can perform 10,000 iterations in 10 hours.

\begin{figure*}[!t]
\centering
\includegraphics[width=\linewidth]{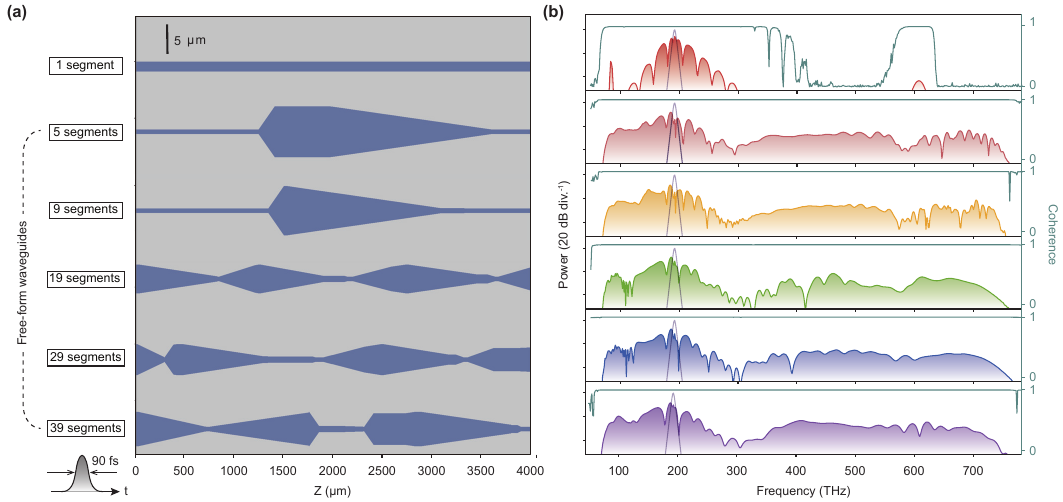}
\caption{{\bf SCG in uniform and free-form waveguides.} (a) Top-view geometries of the waveguides. The scale bar of the vertical axis is to indicate the width of the waveguides. The number of segments for each waveguide is indicated at the input port. (b) Simulated output spectra from the waveguides in (a). The spectrum of the input pulse is indicated in gray. The coherence of the spectra is also plotted in green.}
\label{Fig2}
\end{figure*}

\section{Mechanism}

We first compare the performance of free-form waveguides with uniform waveguides. Here we select Si$_3$N$_4$ waveguides and optimize $FOM$s regarding intensity, flatness, and coherence within both the visible (400-760 nm) and mid-IR (2500-4000 nm) bands. These Si$_3$N$_4$ waveguides are designed with an air cladding on a silica substrate, with a fixed length of 4 mm and a thickness of 740 nm. The input pulse used in our simulation is a sech-shaped pulse with a central wavelength of 1560 nm, peak power of 1.5 kW, and pulse width of 90 fs. Only TE fundamental modes are considered hereafter. The results of SCG are presented in Fig. \ref{Fig2}(a). For the uniform waveguide (1 segment), the produced spectra are relatively narrow and exhibit disjointed coverage in the visible, near-IR, and mid-IR regions. In contrast, the free-form waveguides produce much broader and flatter spectra. For instance, a waveguide with 29 segments generates 700 times and 20 times higher power than the uniform waveguide in the visible and mid-IR bands, respectively. 

We further calculate the first-order coherence $g_{12}^{(1)}$ of the spectra based on
\begin{equation}
g_{12}^{(1)}(\omega)=\frac{\left\langle \tilde{A}_1^{\star}(\omega) \tilde{A}_2(\omega)\right\rangle}{\sqrt{\left\langle\left|\tilde{A}_1(\omega)\right|^2\right\rangle\left\langle\left|\tilde{A}_2(\omega)\right|^2\right\rangle}},
\end{equation}
where $\tilde{A}_1$ and $\tilde{A}_2$ represent two realizations with different random noise seedings \cite{dudley2006supercontinuum}. The averaged coherence is determined using 50 independently simulated spectra. Notably, the SCG in the free-form waveguides manifests near-unity coherence throughout the spectral range under investigation, showing significant advantages over spectra generated from the uniform waveguide.

\begin{figure*}[ht]
\centering
\includegraphics{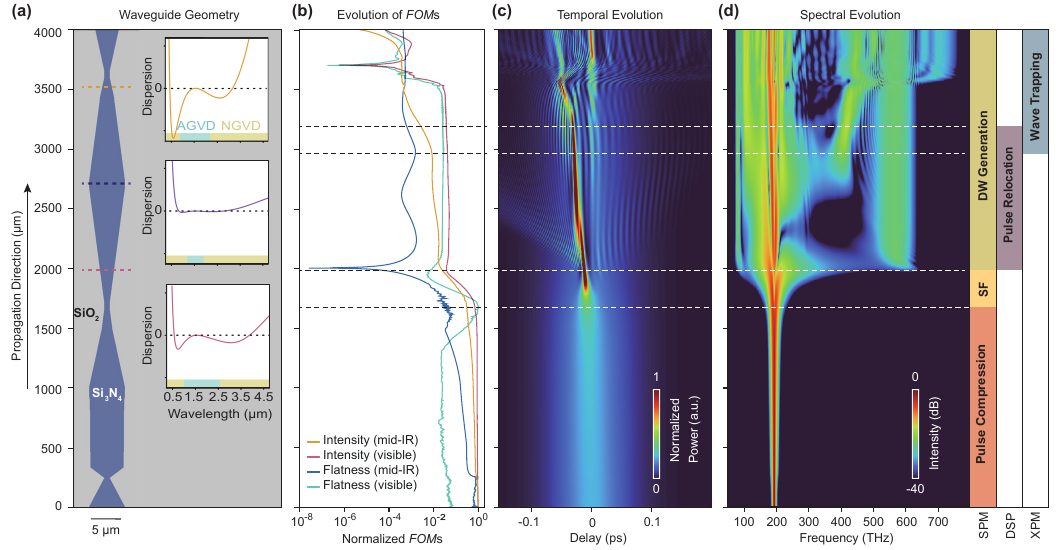}
\caption{{\bf Dynamics of SCG in free-form waveguides.} (a) Top-view of the waveguide geometry. Insets: dispersion profiles at three different locations. AGVD: anomalous group velocity dispersion. NGVD: normal group velocity dispersion. (b) Simulated evolution of $FOM$s. (c) Simulated temporal evolution. (d) Simulated spectral evolution. DW: dispersive wave; SF: soliton fission; SPM: self-phase modulation; DSP: dispersive effect; XPM: cross-phase modulation.
}
\label{Fig3}
\end{figure*}

Although the inverse design is usually considered a black box, understanding the key dynamics that leverage the performance of free-form waveguides of interest. As an example, we investigate SCG in a waveguide consisting of 29 segments (Fig. \ref{Fig3}). Based on the evolution of the $FOM$s, the SCG can be classified into five stages. In the initial stage, the waveguide width gradually changes from 800 nm to 5600 nm over a length of 1.687 mm, leading to pulse compression due to self-phase modulation. Subsequently, the waveguide narrows down to 800 nm, facilitating the generation of a higher-order soliton with a soliton number exceeding 6. In the second stage, soliton fission occurs and significantly broadens the spectrum. Subsequently, dispersive waves start to form in the visible and mid-IR bands. Modulational instability is effectively suppressed due to continuous tuning of the wavelength of the parametric gain through the tapered structure. These phenomena align well with the physical mechanisms commonly employed in the design of coherent SCG \cite{dudley2006supercontinuum}.

However, The insertion of a wide waveguide after soliton fission has minimal impact on the $FOMs$. Indeed, its impacts are more pronounced in the time domain, as the split pulses continue to propagate at speeds determined by their respective group velocities and are relocated at the end of this stage. It ensures substantial temporal overlap between these pulses in the final stage, which further broadens the dispersive waves due to cross-phase modulation. This wave-trapping stage is critical for free-form waveguides to outperform uniform waveguides, and the required precise control of the dispersion to realize pulse relocation remains a challenge for forward design.

\section{Numerical results}

\begin{figure*}[!hbt]
\centering
\includegraphics[width=0.9\linewidth]{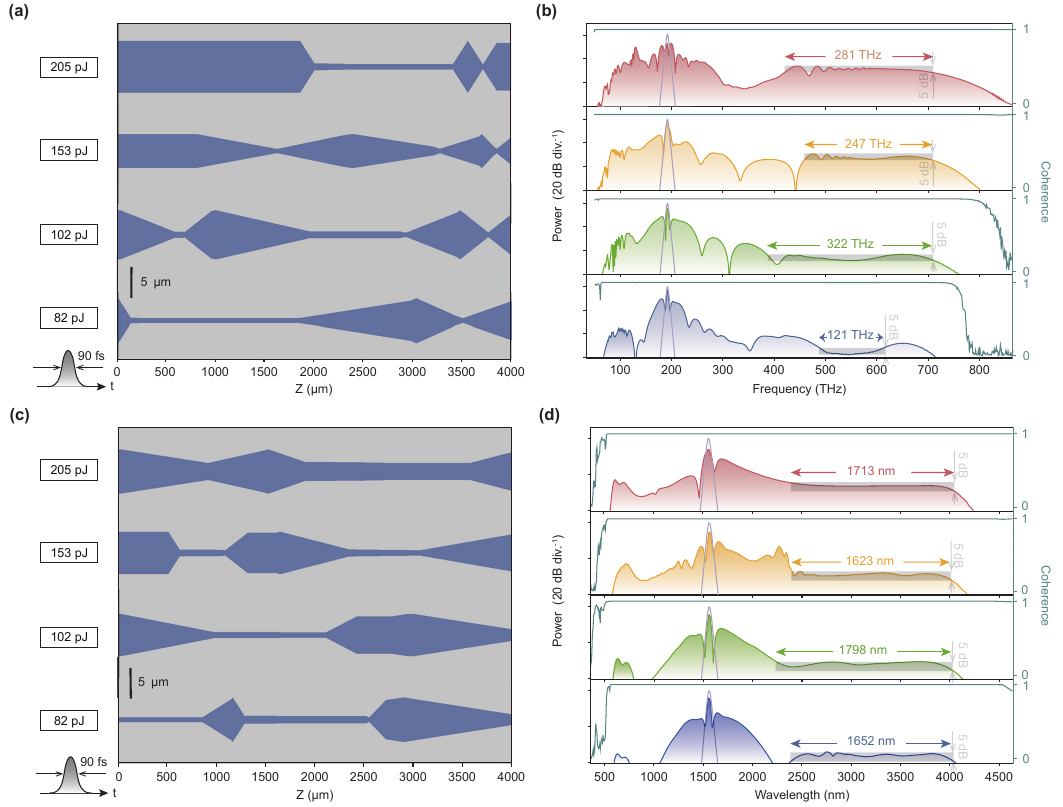}
\caption{{\bf Free-form waveguides for coherent SCG in the visible and mid-IR band.} (a) Waveguide geometries and (b) optical spectra optimized for coherent SCG in the visible band. (c) Waveguide geometries and (d) optical spectra optimized for coherent SCG in the mid-IR band. The input pulse energies are indicated at the input port. The spectra of the input pulse and the coherence are overlaid as gray and green traces, respectively.}
\label{Fig4}
\end{figure*}

In this section, we present some exemplary designs generated by our algorithm. The first example targets broadband visible OFCs within the wavelength range between 400 nm to 760 nm (394 THz to 749 THz). The inverse design is performed for a comb generator operating at 1560 nm with a pulse width of 90 fs and $\rm Si_{3}N_{4}$ waveguides with a length of 4 mm and a thickness of 740 nm (see Fig. \ref{Fig4}(a)). These waveguides are designed for different pulse energies ranging from 82 pJ to 205 pJ. The overall width of the waveguide decreases with lower-power input pulses, indicating a higher nonlinear coefficient required for spectral broadening. The power of the comb lines in the visible band increases with higher input power levels, yet every spectrum is capable of delivering coherent combs spanning a few hundred terahertz with power variations within 5 dB. 

\begin{figure*}[!ht]
\centering
\includegraphics[width=0.9\linewidth]{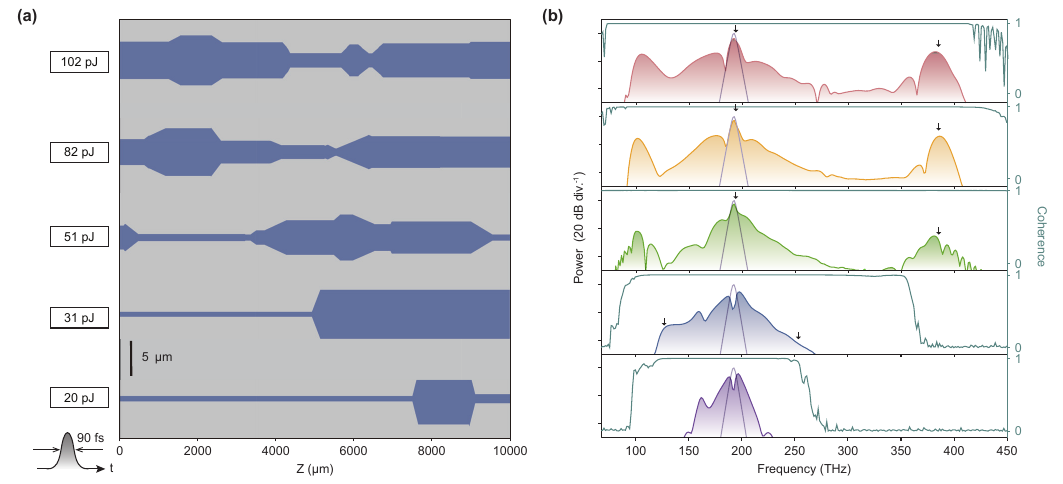}
\caption{{\bf Free-form waveguides for self-referencing of OFCs}. (a) Optimized waveguide geometries for different input pulse energies. The input pulse energies are indicated at the input port. (b) Simulated output spectra from the waveguides in (a). The spectra of the input pulse are indicated in gray. The coherence of the spectra is also plotted in green. Arrows indicate the optimal wavelengths for realizing $f-2f$ self-referencing.}
\label{Fig5}
\end{figure*}

We then perform tests in the mid-IR bands (2500 nm to 4000 nm) with the same settings of the comb generator and waveguide constraints. Excellent coherence and flatness are achieved for all input power levels, with spectral variations below 5 dB achieved over a 1500 nm wavelength span (Fig. \ref{Fig4}(d)). This result has surpassed previous demonstrations in uniform \cite{guo2018mid} and arrayed \cite{guo2020nanophotonic} $\rm Si_{3}N_{4}$ waveguides.

\begin{figure*}[!ht]
\centering
\includegraphics[width=\linewidth]{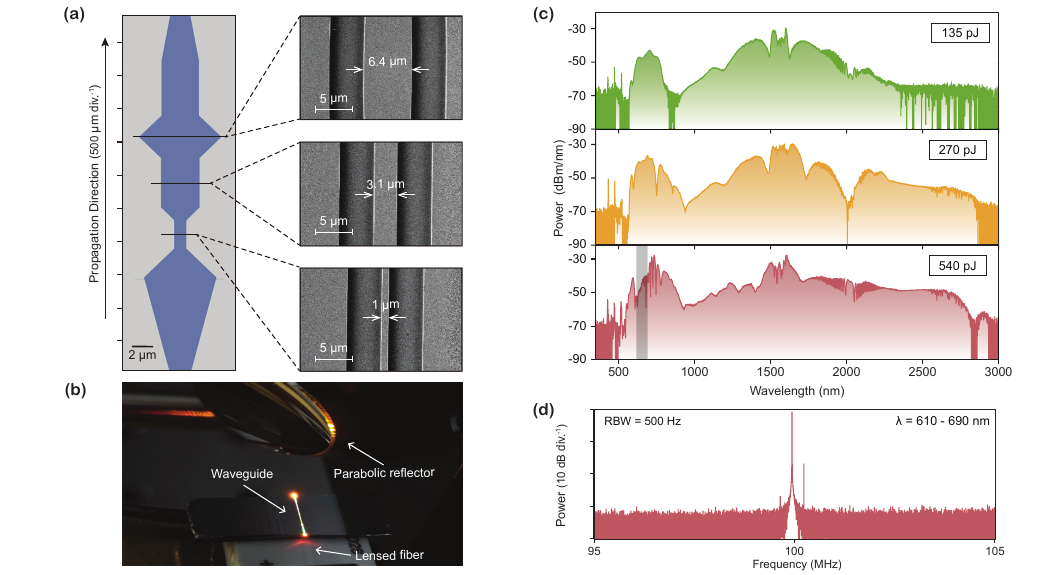}
\caption{{\bf Experimental characterization of SCG in free-form waveguides.} (a) The designed waveguide geometry and the scanning electron microscope images of the fabricated waveguide. (b) Experimental setup for characterization of broadband SCG. (c) Measured optical spectra for input pulse energies of 135 pJ, 270 pJ, and 540 pJ. (d) Measured radiofrequency beatnote of the optical spectra indicated by the shading in the bottom panel of (c). RBW: resolution bandwidth.}
\label{Fig6}
\end{figure*}

Beyond the straightforward optimization problem presented above, our algorithm also allows for the design of waveguides to enable $f-2f$ self-referencing of OFCs. This task does not have a fixed wavelength to optimize, as it only requires the comb to be coherently broadened to an octave, such that the low-frequency ($\omega_o$) components can be doubled to beat with the high-frequency components ($2\omega_o$). The frequency of the beatnote corresponds to the carrier-envelope-offset frequency of the comb. To maximize the signal-to-noise ratio of the beatnote, we first search for the optimal frequency using the following criterion:
\begin{equation}
\omega_o={\mathrm{min}} -\left({\log\left(|A\left(\omega\right)\right|^2)+\log(\left|A\left(\mathrm{2}\omega\right)\right|^2))}\right).
\label{eq:8}
\end{equation}
The $FOM$s for this problem are given by 
\begin{equation}\label{eq:9}
FOM_{1}= -{\log\left(|A\left(\omega_{o}\right)\right|^2)},\\
\end{equation}
\begin{equation}\label{eq:10}
FOM_{2}= \mathrm{-}{\log\left(|A\left(\mathrm{2}\omega_{o}\right)\right|^2)}.\\
\end{equation}
\begin{equation}\label{eq:11}
\begin{split}
FOM_{3}=-\int_{\omega_{o}-\varepsilon{\ }}^{\omega_{o}+\varepsilon{\ }}\left[\log{\left(|A_1\left(\omega\right)\right|^2)}-\log{\left(|A_2\left(\omega\right)\right|^2)}\right]^2\mathrm{d} \omega 
\\
-\int_{\mathrm{2}\omega_{o}-\varepsilon{\ }}^{\mathrm{2}\omega_{o}+\varepsilon{\ }}\left[\log{\left(|A_1\left(\omega\right)\right|^2)}-\log{\left(|A_2\left(\omega\right)\right|^2)}\right]^2\mathrm{d} \omega,
\end{split}
\end{equation}
which respectively measure the intensities of the spectral components and their coherence. $A_1$ and $A_2$ are two different realizations of the spectra with different noise inputs, and $\epsilon$ denotes the phase-matching bandwidth of the doubler. We perform calculations on air-cladded $\rm Si_{3}N_{4}$ waveguides with a fixed length of 10 mm and a thickness of 740 nm, and the results are displayed in Fig. \ref{Fig5}. The optimal frequencies for self-referencing are dynamically adapted to lower pulse energies. For pulse energy above 50 pJ, the lower frequency is chosen to be at the pump frequency of around 200 THz. However, for pulse energy of 31 pJ, no dispersive waves are generated at 400 THz, and the optimal frequencies are shifted to around 125 THz. Note that when the input pulse energy is below 20 pJ, it is not possible to realize octave-spanning OFCs using this comb generator.

\section{Experimental results}

We fabricate $\rm Si_{3}N_{4}$ waveguides to validate the efficacy of our design. The geometry of the 740-nm-thick air–clad $\rm Si_{3}N_{4}$ waveguide is represented in Fig. \ref{Fig6}(a), which has been optimized for generating ultra-broadband OFCs from the visible to mid-IR wavelengths. The patterns are precisely defined through electron-beam lithography and are transferred to the $\rm Si_{3}N_{4}$ layer via reactive ion etching. The input pulse is derived from an erbium-doped-fiber mode-locked laser with a 50 fs pulse duration and a 100-MHz repetition rate. It is coupled to the waveguide through a lensed fiber, with a 5 dB coupling loss per facet. The output from the waveguide is collimated by a parabolic reflector and then coupled into optical fibers by another parabolic reflector (Fig. \ref{Fig6}(b)). This effectively overcomes the chromatic and spherical aberrations associated with free-space optical paths. The spectra shown in Fig. \ref{Fig6}(c) are acquired at three different pulse energy levels, which are analyzed by three optical spectral analyzers operating at different wavelength ranges. For a 135-pJ input pulse, the spectrum extends to an octave, with substantial dispersive waves generated at the visible wavelength. When the pulse energy is increased to 540 pJ, a broad spectrum from 550 nm to 2700 nm emerges with intensity variations within 20 dB. We further filter the spectrum near 650 nm for photodetection. A clean beatnote at 100 MHz is observed (Fig. \ref{Fig6}(d)), which corresponds to the repetition rate of the mode-locked laser, confirming the coherence of the generated spectrum.

\section{Discussion and outlook}

We have demonstrated the inverse design of coherent SCG in free-form nanophotonic waveguides despite nonlinear and stochastic processes. Several future improvements can enhance their advantages over uniform waveguides. First, increasing the length of the waveguide provides a larger parameter space for the design. Indeed, recent advances in fabrication technology have enabled low-loss, meter-long nonlinear waveguides \cite{liu2021high,rebolledo2022coherent}. Second, other nonlinear processes, e.g., Raman and Pockels effect, can contribute to further extending the spectral coverage. For example, LiNbO$_3$-on-insulator and AlN waveguides are capable of generating OFC into the visible or ultra-violet bands via second-order nonlinearity \cite{hickstein2017ultrabroadband,yu2019coherent,lu2019octave,lu2020ultraviolet,chen2021supercontinuum}. Furthermore, the complex spatiotemporal dynamics in multimode waveguides could be harnessed as another degree of freedom \cite{horak2012multimode,krupa2019multimode}. The resulting increased computational load is expected to be resolved by more intelligent algorithms such as machine learning \cite{teugin2020controlling,genty2021machine}. We are also anticipating more functional components to be involved in the design, including periodic gratings \cite{carlson2017photonic}, coupled waveguides \cite{guo2020nanophotonic}, and resonant structures \cite{obrzud2017temporal,anderson2021photonic}. These highly flexible devices have the potential to create diverse light sources for a wide range of applications, such as optical coherence tomography \cite{hartl2001ultrahigh, tu2013coherent, rao2021shot,ji2021millimeter}, ultrashort pulse generation \cite{dudley2004fundamental,heidt2011high}, and optical communications \cite{smirnov2006optical}.

\begin{backmatter}
\bmsection{Funding} This work was supported by National Key R\&D Plan of China (Grant No. 2021YFB2800601), National Natural Science Foundation of China (92150108), Beijing Natural Science Foundation (Z210004), and the High-performance Computing Platform of Peking University.

\bmsection{Disclosures} The authors declare no conflicts of interest.

\bmsection{Data availability} The code and data used to produce the plots within this work will be released on the repository Figshare upon publication of this preprint.

\end{backmatter}

\bibliography{sample}

\bibliographyfullrefs{sample}



\end{document}